\documentclass[amsmath,amssymb,12pt,superscriptaddress,nofootinbib]{revtex4-1}
\usepackage{graphicx}
\usepackage{bm}
\usepackage{color}
\usepackage{soul} 
\usepackage{caption} 
\sethlcolor{yellow}

\newcommand{\dd}{\mathrm{d}}

\definecolor{DarkBlue}{rgb}{0,0,0.7} 

\definecolor{DarkRed}{rgb}{0.65,0,0} 

\definecolor{DarkGreen}{rgb}{0,0.6,0}

\begin{document}
\baselineskip5.5mm
\thispagestyle{empty}

{\baselineskip0pt
\leftline{\baselineskip14pt\sl\vbox to0pt{
               \hbox{\it Division of Particle and Astrophysical Science}
              \hbox{\it Nagoya University}
                             \vss}}
\rightline{\baselineskip16pt\rm\vbox to20pt{
\vss}}%
}

\author{Masato Tokutake}\email{tokutake@gravity.phys.nagoya-u.ac.jp}
\affiliation{ 
Gravity and Particle Cosmology Group,
Division of Particle and Astrophysical Science,
Graduate School of Science, Nagoya University, 
Nagoya 464-8602, Japan
}

\author{Kiyotomo Ichiki}\email{ichiki.kiyotomo@c.mbox.nagoya-u.ac.jp}
\affiliation{
Gravity and Particle Cosmology Group,
Division of Particle and Astrophysical Science,
Graduate School of Science, Nagoya University, 
Nagoya 464-8602, Japan
}

\author{Chul-Moon~Yoo}\email{yoo@gravity.phys.nagoya-u.ac.jp}
\affiliation{
Gravity and Particle Cosmology Group,
Division of Particle and Astrophysical Science,
Graduate School of Science, Nagoya University, 
Nagoya 464-8602, Japan
}

\title{
Observational Constraint on Spherical Inhomogeneity 
\\
with CMB and Local Hubble Parameter 
}

\begin{abstract}
\vskip0.5cm
\baselineskip5.5mm
We derive an observational constraint on a spherical inhomogeneity 
of the void centered at our position from the 
angular power spectrum of the cosmic microwave background(CMB) and 
local measurements of the Hubble parameter. 
The late time behaviour of the void is assumed to be 
well described by the so-called $\Lambda$-Lema\^itre-Tolman-Bondi~($\Lambda$LTB) solution. 
Then, we restrict the models to the asymptotically homogeneous models 
each of which is approximated by a flat Friedmann-Lema\^itre-Robertson-Walker model. 
The late time $\Lambda$LTB models are parametrized by four parameters 
including the value of the cosmological constant and the local Hubble parameter. 
The other two parameters are used to parametrize the observed distance-redshift relation. 
Then, the $\Lambda$LTB models are constructed so that they are compatible with the given 
distance-redshift relation. 
Including conventional parameters for the CMB analysis, 
we characterize our models by seven parameters in total. 
The local Hubble measurements are reflected in the prior distribution of the local Hubble parameter. 
As a result of a Markov-Chains-Monte-Carlo analysis for 
the CMB temperature and polarization anisotropies, 
we found that the inhomogeneous universe models with vanishing cosmological constant are 
ruled out as is expected. 
However, a significant under-density around us is still compatible with 
the angular power spectrum of CMB and the local Hubble parameter. 

\end{abstract}

\maketitle
\pagebreak

\section{Introduction}
\label{intro}

In observational cosmology, 
the global homogeneity and isotropy is a commonly unquestioned hypothesis, 
which is therefore called the cosmological principle. 
Actually, homogeneous and isotropic universe models have achieved great success 
to explain observational data and describe our universe. 
Nevertheless, it is interesting to ask 
how large magnitude of cosmological scale inhomogeneity can be 
compatible with the current cosmological observations. 
The observational test of the cosmological principle may be 
one of the most fundamental issues in cosmology just like old times. 
From this viewpoint, here 
we consider an observational constraint on cosmological scale inhomogeneity 
with the cosmic microwave background~(CMB) and the local Hubble parameter. 
Since the isotropy of the universe is strongly supported by the 
isotropy of the CMB temperature, 
we focus on spherically symmetric inhomogeneous universe models. 

Once we are allowed to be at the center of the universe 
with a spherical inhomogeneity, 
since most observables are limited on our past lightcone, 
the spatial inhomogeneity and temporal dependence 
may degenerate with each other. 
This fact gives one of the main difficulties in analyses of 
inhomogeneous universe models differently from 
homogeneous and isotropic universe models. 
Therefore, careful evaluation of observables 
and multi-directional analyses are important for the 
observational test of spherical inhomogeneity of our universe. 
We make a contribution to this issue from one direction in this paper.

Spherically symmetric dust universe models, 
so called the Lema\^itre-Tolman-Bondi (LTB) 
models~\cite{Lemaitre:1933gd,Tolman:1934za,Bondi:1947av}, 
have been extensively studied in the last decade. 
The LTB models have been attracted much attention mainly as 
an alternative scenario to explain the apparent accelerated expansion 
of our universe without dark energy~\cite{Zehavi:1998gz,Celerier:1999hp,Tomita:1999qn}. 
Actually, it is known that there exist the LTB models which 
can explain the observed luminosity distance redshift relation without 
a cosmological constant 
$\Lambda$ \cite{Celerier:1999hp,Iguchi:2001sq,Yoo:2008su}. 
Especially, void-type inhomogeneity composed of growing modes 
has been actively studied because it can be compatible with 
the inflationary paradigm and the apparent accelerated expansion. 
Eventually, it has been revealed that 
the apparent accelerated expansion cannot be explained only by 
the radial inhomogeneity without a cosmological constant 
if we assume the standard cosmological history 
before the last scattering surface of 
CMB photons~(see, e.g. Refs.~\cite{Redlich:2014gga,Valkenburg:2012td} for a detailed analysis). 

In contrast, the models with a non-vanishing cosmological constant, namely, 
$\Lambda$LTB models, have been studied not very often 
although we can find several restricted 
analyses~\cite{Marra:2010pg,Valkenburg:2012td,Negishi:2015oga,Ichiki:2015gia}. 
In this paper, 
we assume that the late time behaviour of our universe is well described by 
a $\Lambda$LTB model. 
One remarkable feature in our approach is 
that we specify the spherically symmetric inhomogeneity 
by using the so-called inverse construction from a distance redshift relation. 
The same procedure is adopted in Ref.~\cite{Sundell:2015cza}. 
The inverse construction is a method to construct the $\Lambda$LTB model 
in which the distance redshift relation for the central observer agrees with 
the designated one. 
In the case of $\Lambda$LTB models, as is shown in Ref.~\cite{Tokutake:2016hod}, 
once the value of the cosmological constant is fixed and 
the angular diameter distance is specified as a function of 
the redshift, we can uniquely determine the $\Lambda$LTB model. 
Therefore, the parameters to specify a $\Lambda$LTB model 
are equivalent to the parameters contained in the distance redshift relation 
beside the value of the cosmological constant. 
In our analysis, the distance redshift relation is assumed to be 
given by the same form as the distance 
in the dust-dominated Friedmann-Lema\^itre-Robertson-Walker~(FLRW) universe
and parametrized by the Hubble constant $H_0$ and 
two fictitious cosmological parameters 
$\Omega^{\rm dis}_{\rm m0}$ and $\Omega^{\rm dis}_{\rm \Lambda0}$. 
It should be emphasized that 
$\Omega^{\rm dis}_{\rm m0}$ and $\Omega^{\rm dis}_{\rm \Lambda0}$ 
are not necessarily related to the real matter density and 
the cosmological constant $\Lambda$ 
but just parameters to specify a distance redshift relation. 
This procedure is different from conventional 
methods of artificial direct parametrization of LTB models 
adopted in previous works~(see, e.g. \cite{GarciaBellido:2008nz,Valkenburg:2012td}). 
Therefore, it could be possible to extract unknown effects of 
the spherical inhomogeneity around us.

In order to keep the predictability of the CMB anisotropy, 
we restrict our attention to asymptotically homogeneous models, 
and gradually connect each of the $\Lambda$LTB models to 
a flat FLRW universe model. 
The connection is performed in the redshift interval $2< z<15$, 
and the models are described by the standard homogeneous and isotropic universe models 
including relativistic energy components before $z=15$. 
The late time $\Lambda$LTB models are parametrized by four parameters 
including the value of the cosmological constant and the local Hubble parameter. 
Including conventional parameters for the CMB analysis, 
we characterize our models by seven parameters in total. 
For these seven parameters, we perform a 
Markov Chain Monte Carlo~(MCMC) analysis by modifying 
the package CosmoMC\footnote{http://cosmologist.info/cosmomc/readme.html}. 
The local Hubble measurements are reflected in the prior distribution of the local 
Hubble parameter.

This paper is organized as follows. 
In Sec.~\ref{section model}, 
we briefly review the inverse construction method reported 
in Ref.~\cite{Tokutake:2016hod} 
and how to construct a universe model in the late time domain. 
The method to calculate the CMB angular power spectrum 
and the parameter set for the MCMC analysis is summarized in Sec.~\ref{CMBMCMC}. 
In Sec.~\ref{section results}, we show contour maps 
of the allowed regions for substantial parameters including 
the amplitude of the under-density. 
Sec.~\ref{section summary and discussion} is devoted to a summary and discussion.

In this paper, we use geometrized units in which 
the speed of light and Newton's gravitational constant are one, 
respectively.

\section{Late time model construction}
\label{section model}

The LTB solution is the solution for the Einstein equations 
of the spherically symmetric dust fluid system. 
A line element of the LTB solution is written in the form:
\begin{equation}
\dd s^2 = -\dd t^2+\frac{(\partial_r R(t,r))^2}{1-k(r)r^2}\dd r^2 + R^2(t,r)\dd \Omega^2, 
\end{equation}
where $R(t,r)$ is the areal radius and $k(r)$ is the function of 
the radial coordinate $r$ called the curvature function. 
From the Einstein equations with the cosmological constant $\Lambda$, 
we obtain the following equation:
\begin{equation}
(\partial_tR)^2 = -k(r)r^2+\frac{m(r)}{3R}+\frac{1}{3} \Lambda R^2 := f(r,R), 
\label{Eeq}
\end{equation}
where $m(r)$ is an arbitrary function of $r$. 
The comoving energy density $\rho$ is given by 
\begin{equation}
\rho(t,r)=\frac{1}{4\pi}\frac{\partial_rM(r)}{R^2\partial_rR}
\end{equation}
with $M(r)=m(r)r^3/6$. 
We can formally integrate Eq.~\eqref{Eeq} as
\begin{equation}
t-t_B(r) = \int_{0}^{R}\frac{dX}{\sqrt{f(r,X)}},
\label{Eeqsol}
\end{equation}
where $t_B(r)$ is the function of $r$ which gives the bigbang time. 
The LTB solution has three arbitrary functions $k(r)$, $m(r)$ and $t_B(r)$. 
Since the inhomogeneity associated with $t_{\rm B}(r)$ 
corresponds to decaying modes, 
we simply assume $t_B = \rm{const.}$ in this paper, 
where the constant value can be set to zero by shifting the origin of the time.
In addition, by using the gauge degree of freedom to choose the radial coordinate $r$, 
we set $m={\rm const.}$. 

In Ref.~\cite{Tokutake:2016hod}, 
it is shown that, for $t_B = \rm{const}$, 
$k(r)$ and the value of $m$ are uniquely determined, 
once the Hubble parameter $H_0$ and 
the normalized cosmological constant $\Omega_{\Lambda 0}$ are fixed and 
the cosmological distance $D(z)$ is given as a function of the redshift $z$. 
In this paper, we use the same functional form of $D(z)$ as that in the 
matter dominated homogeneous and isotropic universe models:
\begin{equation}
D(z) = D_{\Lambda \rm{CDM}}(z;\Omega_{m0}^{\rm{dis}},\;\Omega_{\Lambda 0}^{\rm{dis}},\;H_0),
\end{equation}
where $\Omega^{\rm dis}_{m 0}$ and $\Omega^{\rm dis}_{\Lambda 0}$ are 
the normalized matter density and the cosmological constant for 
the reference homogeneous and isotropic universe. 
It should be noted that $\Omega^{\rm dis}_{m 0}$ and $\Omega^{\rm dis}_{\Lambda 0}$ 
are not necessarily related to the real matter density 
and the cosmological constant $\Lambda$ but just parameters to specify the distance-redshift relation. 
For later convenience, we define $\mathcal{R}_{\Lambda}$ as 
\begin{equation}
\mathcal{R}_{\Lambda} := \frac{\Omega_{\Lambda 0}}{\Omega_{\Lambda 0}^{\rm{dis}}},  
\end{equation}
where $\Omega_{\Lambda 0}:=\Lambda/(3H_0^2)$. 
Then, the $\Lambda$LTB models are parametrized by 
the four parameters:$H_0$, $\mathcal{R}_{\Lambda}$, 
$\Omega^{\rm dis}_{m 0}$ and $\Omega^{\rm dis}_{\Lambda 0}$. 
Readers may refer to Ref.~\cite{Tokutake:2016hod} for details of the construction method. 
Here we note that, by using above procedure, we can obtain 
the curvature function $k$ and the redshift $z$ as functions of 
the radial coordinate $r$.

\begin{figure}[htbp]
\begin{center}
\includegraphics[scale=0.8]{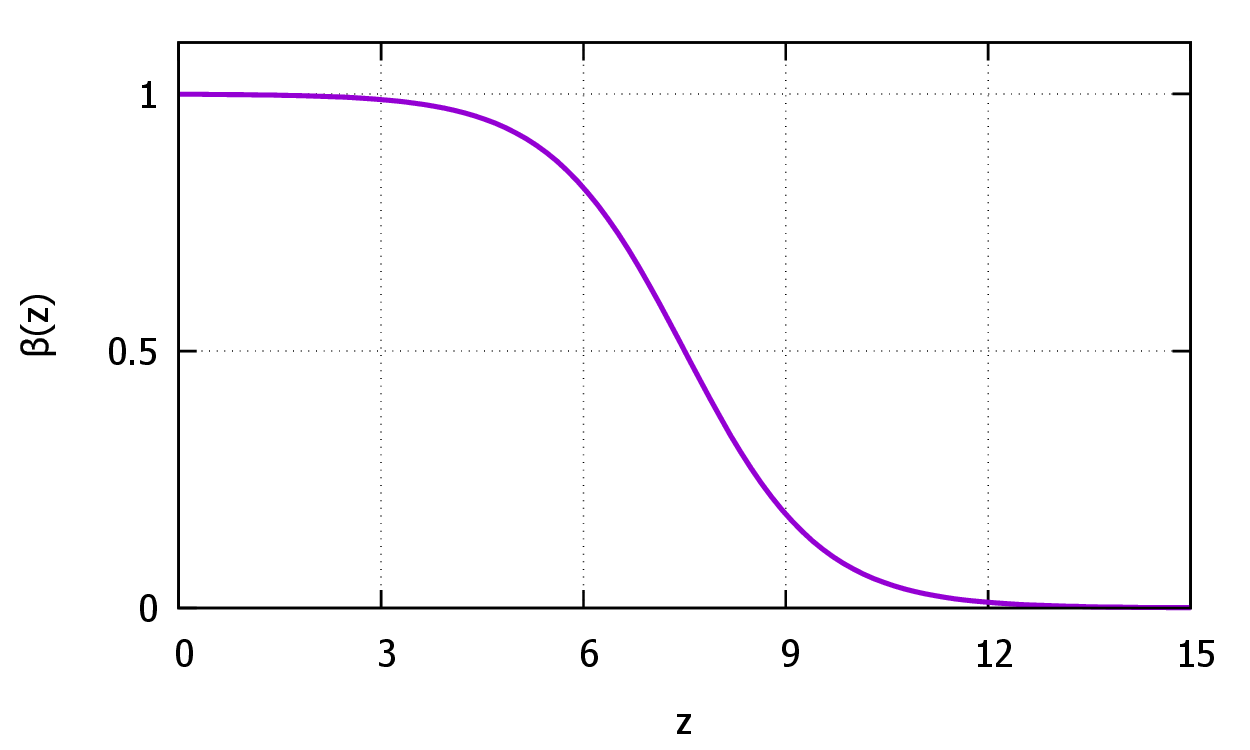}
\caption{Functional form of $\beta(z)$. 
}
\label{fig:beta}
\end{center}
\end{figure}

\begin{figure}[htbp]
\begin{center}
\includegraphics[scale=1.3]{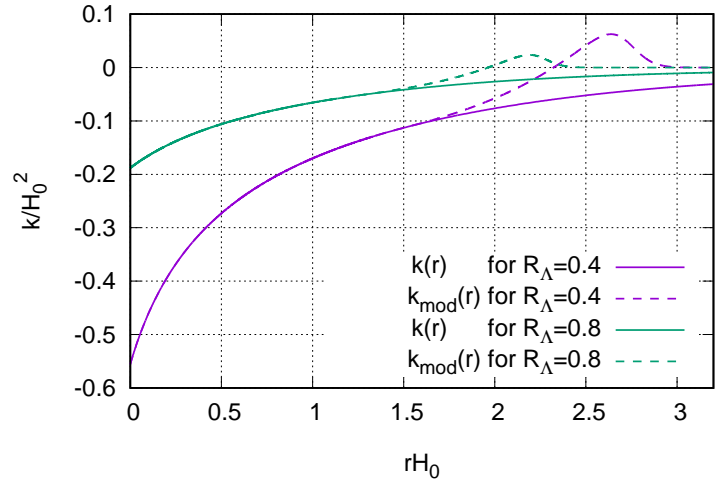}
\caption{Curvature functions $k(r)$ and $k_{\rm mod}(r)$ for $\Omega^{\rm dis}_{m 0}=0.3$ 
and $\Omega^{\rm dis}_{\Lambda 0}=0.7$, and $\mathcal{R}_{\Lambda}=0.4$ and $0.8$. }
\label{fig:kplot}
\end{center}
\end{figure}

As is mentioned in Sec.~\ref{intro}, 
we focus on the models each of which asymptotically coincides with a flat FLRW model. 
For this purpose, we gradually connect each LTB model to a 
flat FLRW universe model through the redshift domain $2<z<15$, where 
we do not have significant observational constraints. 
Specifically, for $z>2$, we assume the following form of 
the curvature function $k$:
\begin{equation}
k(r) \rightarrow k_{\rm mod}(r)=\beta(z(r)) 
\biggr[k_{z=2}+\left(\frac{\dd k}{\dd z}\right)_{z=2}(z(r) - 2) 
\biggr],
\label{function form of k}
\end{equation}
where $z(r)$ is the redshift as a function of $r$ 
given for a $\Lambda$LTB model specified by the four parameters:$H_0$, $\mathcal{R}_{\Lambda}$, 
$\Omega^{\rm dis}_{m 0}$ and $\Omega^{\rm dis}_{\Lambda 0}$, %
and 
\begin{equation}
\beta(z) = \frac{1}{1+\exp[3(z-z_b/2)/2]} 
\end{equation}
with $z_b=15$~(see Fig.~\ref{fig:beta} for the functional form of $\beta(z)$). 
Examples of the curvature function $k(r)$ and $k_{\rm mod}(r)$ are shown 
in Fig.~\ref{fig:kplot}. 
%

\begin{figure}[htbp]
	\begin{center}
		\includegraphics[scale=1]{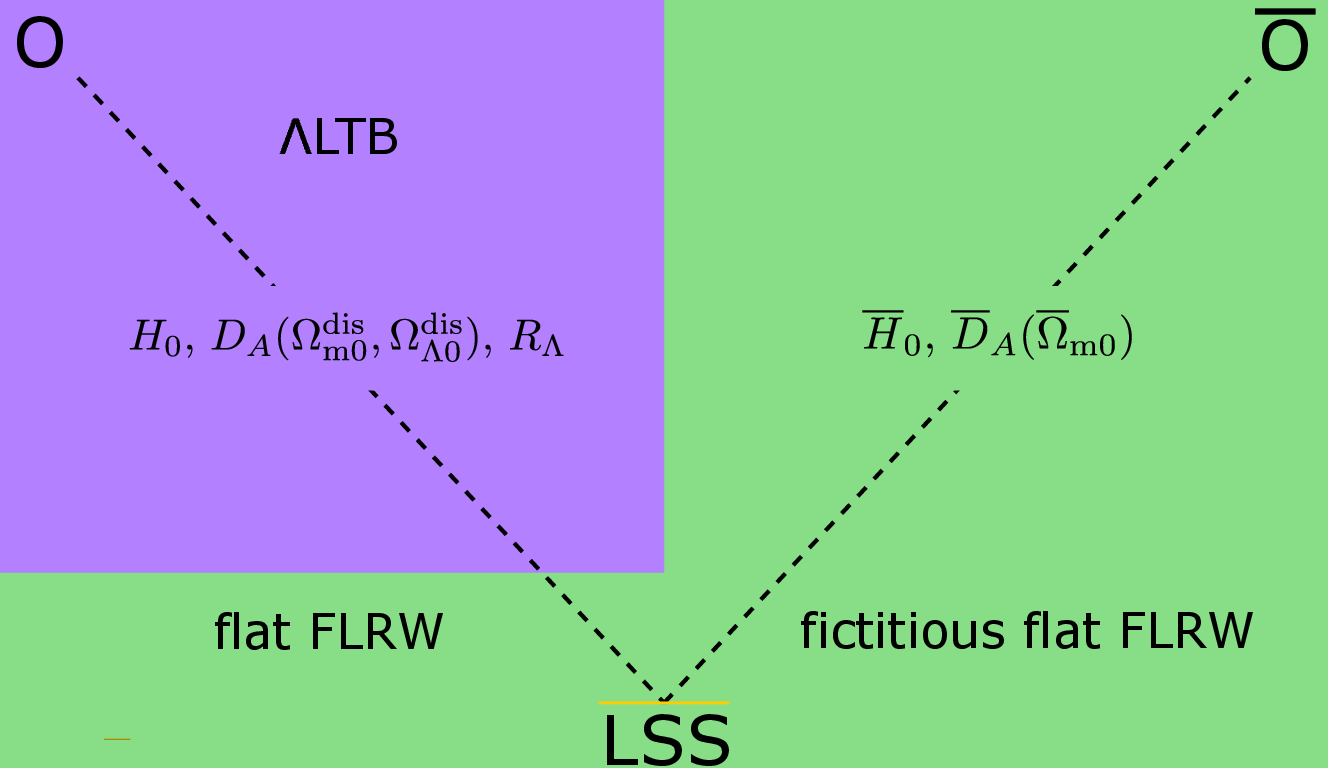}
		\caption{
		A schematic figure for the universe models. 
		}
		\label{our model}
	\end{center}
\end{figure}

In the higher redshift region, for accurate calculation of the CMB spectrum, 
we need to describe our universe taking the 
contribution of radiation components into account~(Fig.~\ref{our model}). 
In this paper, 
we simply add the radiation components with 
the density
$\rho_{r\rm{b}} = 7.804 \times 10^{-34}\left(1+z_{\rm b}\right)^4{\rm g ~cm^{-3}}$ 
at $z_{\rm{b}}$. 
Since the radiation effect 
for the dynamics of the universe is negligible for $z<z_{\rm b}=15$, 
the gap of the Hubble expansion rate at $z=z_b$ is negligible. 
Then the discontinuity of this procedure 
does not significantly affect the final results. 
Actually, we have confirmed that the results 
do not depend on the value of $z_{\rm{b}}$
for $2<z_{\rm b}<15$. 

\section{CMB anisotropy and the MCMC analysis}
\label{CMBMCMC}

\subsection{Angular power spectrum}
\label{section angular power spectrum}

For the calculation of the CMB temperature anisotropy, 
we use the open code CAMB \footnote{http://camb.info}. 
Since homogeneous and isotropic universe models are supposed in this 
code, we need to appropriately modify input parameters 
and the output temperature anisotropy for our purpose. 
In this paper, we mainly focus on the primary effects on 
the CMB anisotropy, that is, 
we consider the temperature anisotropy that originates from 
inhomogeneity of the gravitational potential on the LSS. 
Inhomogeneity in our models are composed of growing modes, 
and it may significantly affect the secondary effects on 
the CMB anisotropy in low $\ell$ domain. 
Therefore, in our analysis, 
we simply ignore the angular power spectrum $C_l$ for $\ell<20$. 

In order to calculate $C_\ell$ observed at the center, 
we consider the fictitious flat FLRW universe which shares the same 
LSS with the inhomogeneous universe of our interest~(see Fig.~\ref{our model}). 
We define the cosmological parameters, the angular diameter distance to LSS and the 
angular power spectrum in the fictitious FLRW model as $\overline{\Omega}_{X0}$, 
$\overline{D}_{\rm A}^{\rm lss}$ and $\overline{C}_{\overline{\ell}}$, respectively. 
The cosmological parameters are fixed when we connect a late time LTB universe model 
to the corresponding flat FLRW universe model at $z=z_{\rm b}$. 
Then, the primary effects on the temperature anisotropy are 
given by those in the fictitious FLRW model, 
while the late time behaviour of the universe model is different from the homogeneous universe.  
Therefore we need to take the difference of the angular diameter distance 
between these models into account. 
Since we focus on $\ell > 20$, in this range, the flat-sky approximation is valid. 
In the flat-sky approximation, 
$C_{\ell}$ is given by the following form~(see, e.g. \cite{2008cmb..book.....D}) 
\begin{equation}
C_\ell = \left( \frac{\overline{D}_{\rm A}^{\rm lss}}{D_{\rm A}^{\rm lss}} \right)^2 
\overline{C}_{\bar\ell}, \label{correctionda}
\end{equation}
where $\bar\ell=\ell ~\overline{D}_{\rm A}^{\rm lss}/D_{\rm A}^{\rm lss}$. 
The flat-sky approximation is valid in the accuracy of 1\% for $l > 20$ \cite{Vonlanthen:2010cd}. 

\subsection{Parameter set for MCMC and calculation of $C_\ell$}
\label{section MCMC}
In order to describe the CMB anisotropy, in addition to the cosmological parameters 
for the late time universe model, we introduce the following three parameters: 
the scalar spectral index $n_s$, 
amplitude of primordial fluctuation $A$ 
and baryon to matter ratio $\alpha:=\overline{\Omega_{b0}}/\overline{\Omega_{m0}}$. 
In summary, we have the following seven free parameters:\{$\Omega_{m0}^{\rm{dis}}$, $\Omega_{\Lambda 0}^{\rm{dis}}$, $H_0$, $\mathcal{R}_\Lambda$, $n_s$, $A$, $\alpha$\}. 
In our analysis, we fix the optical depth as $\tau = 0.1$ 
because the analysis does not use the low $\ell$ angular power spectrum, 
in which $\tau$ dependence is significant. 
We set a prior distribution for the Hubble parameter $H_0$ 
which is consistent with the observational 
value given in Ref.~\cite{Efstathiou:2013via}. 
That is, we restrict the value of $H_0$ by using the Gaussian prior 
with $72.5 \pm 2.5\;[\rm{km\;s^{-1}\;Mpc^{-1}}]$, where the interval 
is the $1\sigma$ range. 

\begin{figure}[htbp]
	\begin{center}
		\includegraphics[scale=1]{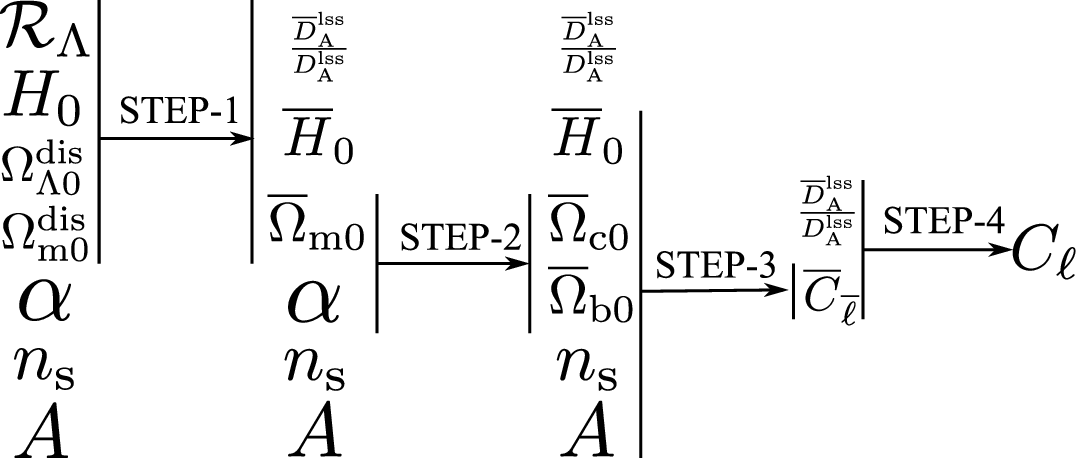}
		\caption{Parameter set for MCMC method and procedure to calculate $C_\ell$. }
		\label{parameterization}
	\end{center}
\end{figure}

Our procedure to calculate $C_\ell$ is summarized in Fig.~\ref{parameterization}. 
The calculation can be summarized in the following 4 steps. 
\begin{itemize}
\item{\bf Step 1}

First, we solve the inverse problem to obtain the $\Lambda$LTB model in $z\leq2$ 
as is discussed in Sec.~\ref{section model} 
by using the four parameters:\{$\Omega_{m0}^{\rm{dis}},
\;\Omega_{\Lambda0}^{\rm{dis}},\;\mathcal{R}_\Lambda,\;H_0$\}. 
For $2<z\leq15$, we give the functional form of $k(z)$ by 
Eq.~\eqref{function form of k}. 
Then, we can calculate $D_A$ and 
determine $\overline{\Omega}_{m0}$, $\overline{H}_0$ 
and $\overline{D}_{\rm A}^{\rm lss}$. 

\item{\bf Step 2}

Second, we fix the normalized baryon density $\overline{\Omega}_{b0}$ and the dark matter density 
$\overline{\Omega}_{c0}$ in the fictitious FLRW model as follows: 
\begin{eqnarray}
\overline{\Omega_{b0}} &=& \alpha\overline{\Omega_{m0}} , \\
\overline{\Omega_{c0}} &=& \overline{\Omega_{m0}} - \overline{\Omega_{b0}}. 
\end{eqnarray}

\item{\bf Step 3}

Then, we calculate $\overline{C}_{\overline{l}}$ which is the angular power spectrum observed at $z=0$ in the fictitious flat FLRW model by inputting the parameters: $\overline{\Omega_{b0}},\;\overline{\Omega_{c0}},\;\overline{H_0},\;n_s$ and $A$ to CAMB. 

\item{\bf Step 4}

Finally, we perform the correction given in Eq.~\eqref{correctionda} to get $C_l$. 
\end{itemize}

\label{section parameter}

Because of the specification of CosmoMC, 
in the actual analysis, 
$\Omega_{\rm m0}^{\rm{dis}}$, $\Omega_{\Lambda0}^{\rm{dis}}$, $H_0$ and $\alpha=\Omega_{\rm b0}^{\rm{dis}}/\Omega_{\rm m0}^{\rm{dis}}$ are derived from another four parameters: 
$\Omega_{b0}^{\rm{dis}}h^2$, $\Omega_{c0}^{\rm{dis}}h^2$, $\Omega_{K0}^{\rm{dis}}$ and the ratio $\theta$ between the sound horizon and the angular diameter distance, where $\Omega_{\rm b0}^{\rm{dis}} + \Omega_{\rm c0}^{\rm{dis}} = \Omega_{\rm m0}^{\rm{dis}}$ and $h$ is the dimensionless Hubble parameter defined as $h = H_0/100$. 

\section{Results}
\label{section results}

\begin{figure}[htbp]
	\begin{center}
		\hspace{-1.5cm}\includegraphics[scale=0.65]{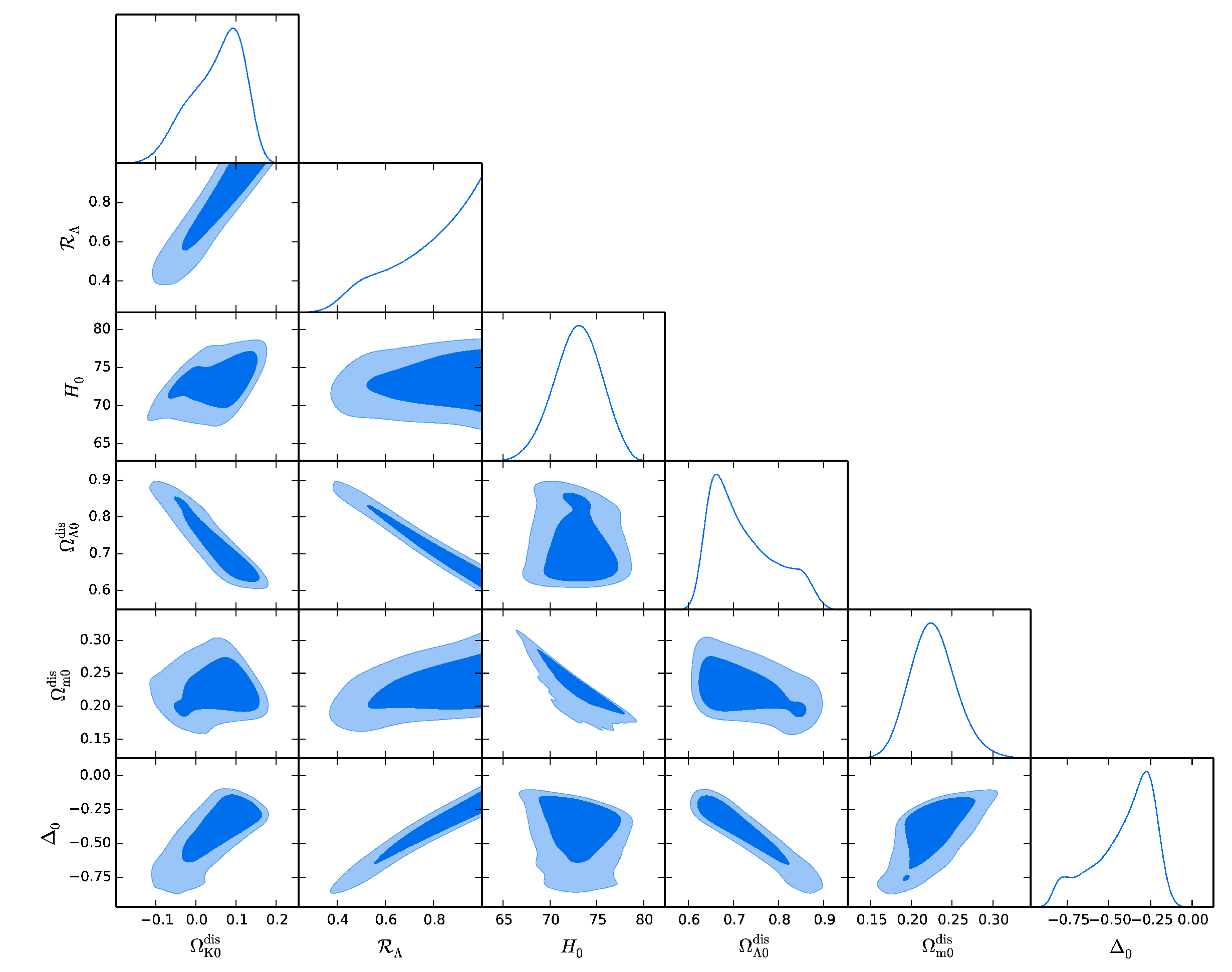}
		\caption{The posterior probability distribution and contour map for the main parameters. 
The dark blue region represents the restriction of $1\sigma$ confidence level and the watery blue region is 
$2\sigma$ confidence level. }
		\label{results1}
	\end{center}
\end{figure}

First, we show posterior distributions and contour maps of $\Omega_{K0}^{\rm{dis}}:=1-\Omega_{\Lambda 0}^{\rm dis}-\Omega_{\rm m0}^{\rm{dis}}$, $\mathcal{R}_\Lambda$, $H_0$, $\Omega_{\Lambda 0}^{\rm dis}$, $\Omega_{\rm m0}^{\rm{dis}}$ 
and $\Delta_0$ in Fig.~\ref{results1}, 
where $\Delta_0$ is the void depth defined by 
\begin{equation}
\Delta_0:=\frac{\rho(t_0,0)-\rho(t_0,r_{\rm lc}(z_{\rm b}))}
{\rho(t_0,r_{\rm lc}(z_{\rm b}))}, 
\end{equation}
with $t_0$ being the initial time satisfying $\partial_tR/R=H_0$ at the center. 
The value of $\mathcal{R}_\Lambda$ is closely correlated with the void depth $\Delta_0$. 
As is shown in this figure, 
$\mathcal{R}_\Lambda$ is restricted $\mathcal R_\Lambda>0.4$ at $2\sigma$ confidence level.  
This result explicitly shows the exclusion of $\Lambda=0$ void models. 
It should be noted that our prior models 
include the flat FLRW models with positive values of $\Lambda$ 
and also the inhomogeneous universe models with $\Lambda=0$ differently from the previous works
not including $\Lambda$. 
Therefore the comparison can be done within the common parameter space, 
and the exclusion is more explicit(see also Ref.~\cite{Valkenburg:2012td}).

\begin{figure}[htbp]
	\begin{center}
\hspace{-4cm}
		\includegraphics[width=10cm]{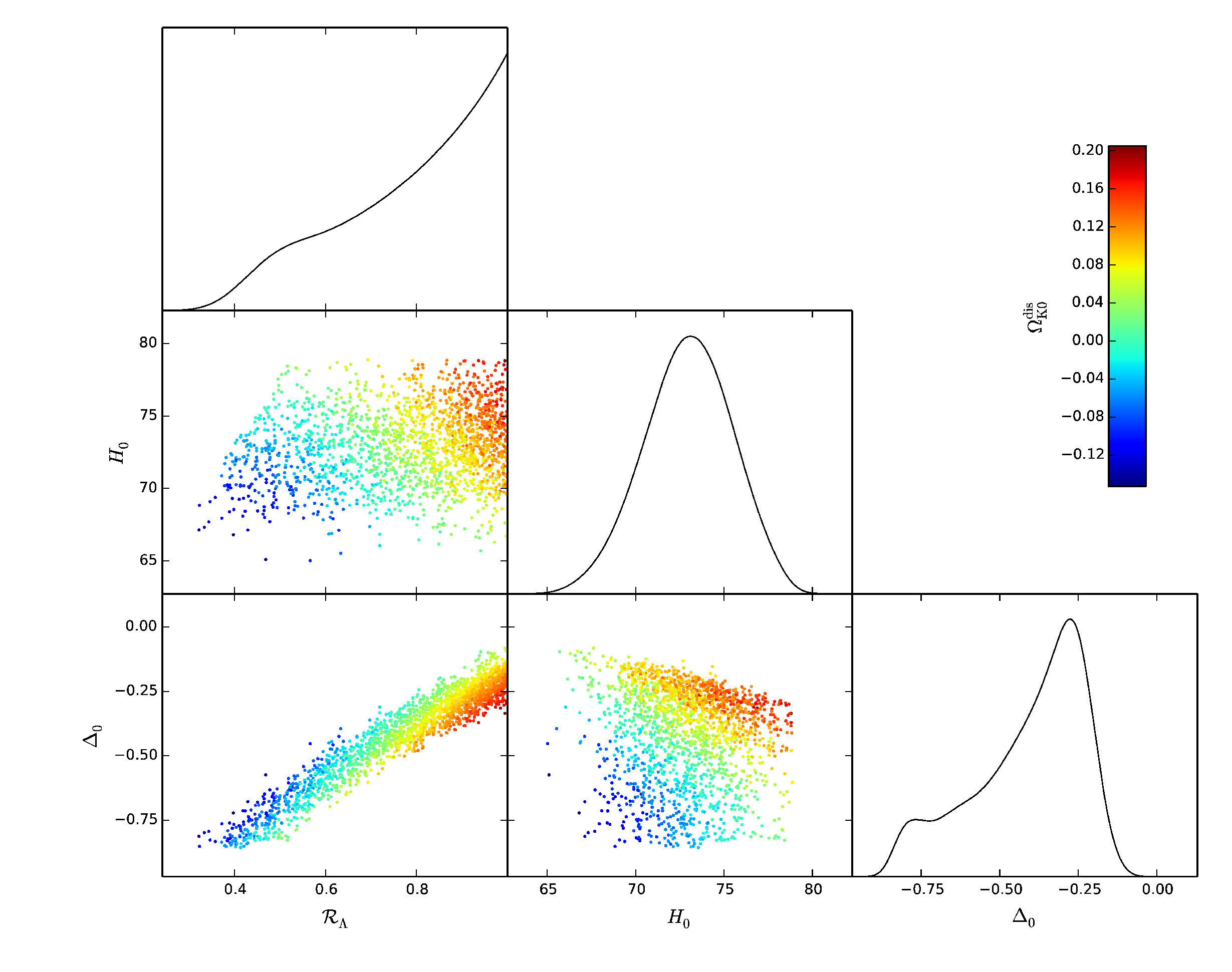}
		\caption{The dependence of $\mathcal{R}_\Lambda,\; H_0$ and void depth $\Delta_0$ with the color plot by the value of $\Omega_{K0}^{\rm{dis}}$.}
		\label{results2}
	\end{center}
\end{figure}

We show the posterior distribution of the void depth $\Delta_0$, 
$\mathcal{R}_\Lambda$ and $H_0$, with  
$\Omega_{K0}^{\rm{dis}}$ dependence by the color plot in Fig.~\ref{results2}. 
Figure~\ref{results2} shows that 
the smaller value of $\mathcal R_{\rm \Lambda}$ implies the larger value of 
the void depth $\Delta_0$. 
Once we fix the value of $\Omega^{\rm dis}_{\rm K0}$, 
$\Delta_0$ tends to be smaller~(deeper void) for the larger value of $H_0$. 
This observation is consistent with 
the previous works~\cite{Marra:2013rba,Ichiki:2015gia}. 
This dependence can be roughly understood as follows. 
If we increase the value of $H_0$, the distance to the LSS decreases. 
On the other hand, we can increase the distance by making void depth deeper 
because the central region becomes closer to an open universe. 
Therefore, the correlation between $H_0$ and $\Delta_0$ 
comes from the compensation of the distance to the LSS. 

It is commonly expected that an under-dense region tends to increase the value of the local Hubble parameter $H_0$ 
compared to the asymptotic value given by CMB observations. 
However, in Fig.~\ref{results1}, the correlation between $H_0$ and $\Delta_0$ is not clear. 
As is mentioned above, the reason for this behaviour comes from $\Omega^{\rm dis}_{\rm K0}$ dependence of $H_0$. 
Even if the value of $\Delta_0$ is fixed at some value, 
changing the value of $\Omega^{\rm dis}_{\rm K0}$, we obtain a different profile of inhomogeneity and 
find a different value of $H_0$. 
This dependence might help the resolution of the $H_0$ tension(see Ref.~\cite{Riess:2016jrr} 
for a recent analysis of the local $H_0$, and Ref.~\cite{Bernal:2016gxb} for possible explanations about the $H_0$ tension).

\section{Summary and discussion}
\label{section summary and discussion}

We have discussed an observational constraint on the spherically symmetric inhomogeneous models 
by the CMB angular power spectrum and local Hubble parameter. 
We assumed that the late time cosmological models are well described by 
$\Lambda$LTB models each of which is characterized by two parameters in the distance-redshift relation, 
the value of the cosmological constant $\Lambda$ and the local Hubble parameter $H_0$. 
Connecting each of the late time inhomogeneous models to a flat homogeneous universe model, 
we calculated the CMB power spectrum observed at the center. 
The MCMC analysis with the Planck data~\cite{Ade:2015xua} explicitly excluded inhomogeneous models with $\Lambda=0$. 
However, at the same time, our results show that 
a significant amplitude of the under-density can be still compatible with the CMB angular power spectrum and 
the local Hubble measurement. 
We found that, even if we fix the amplitude of the void, 
the value of local Hubble parameter can change depending on the parameter $\Omega^{\rm dis}_{\rm K0}$, 
which specifies the inhomogeneity. 
This dependence could help to resolve the $H_0$ tension between the local measurement and 
CMB observations.

Finally, we list related important issues which we could not address in this paper. 
In Ref.~\cite{Valkenburg:2012td}, the strongest constraint for the amplitude of the inhomogeneity 
comes from the linear kinetic Snyaev-Zeldovich effect on the CMB power spectrum in large scales, 
and they concluded that the void amplitude $|\delta_0|$ is smaller than 0.29(see also Refs.~\cite{GarciaBellido:2008gd,Yoo:2010ad,Zhang:2010fa,Moss:2011ze,Bull:2011wi,Ade:2013opi}). 
The Planck team reported the constraint on the kSZ monopole as $72\pm60{\rm km}$ for $z<1$ from the cluster SZ effect. 
This constraint may give much more stringent constraints on the void depth. 
The difference between the radial and transverse BAO scale may be also very efficient indicator for the 
spherical inhomogeneity(see, e.g., Refs.~\cite{Biswas:2010xm,GarciaBellido:2008yq,Zumalacarregui:2012pq,Clarkson:2012bg}).
In this paper, we ignored the angular power spectrum for lower multipoles. 
The spherical inhomogeneity may enhance the Integrated Sachs Wolfe(ISW) effect in the low multipoles. 
In order to clarify the significance of the spherical inhomogeneity to the ISW effect, 
we need to calculate the evolution of the perturbation with spherical inhomogeneity. 
The calculation of the perturbation is also needed for the calculation of the CMB lensing. 
We leave all these issues as future works. 

\section*{Acknowledgements}
This work was supported by JSPS
KAKENHI Grant Numbers JP16K17688, JP16H01097 (CY) and JP16H01543 (KI).


\end{document}